\documentclass{article}

\usepackage[english]{babel}
\usepackage{authblk}
\usepackage{amsfonts}
\usepackage[letterpaper,top=2cm,bottom=2cm,left=3cm,right=3cm,marginparwidth=1.75cm]{geometry}

\usepackage{amsmath}
\usepackage{graphicx}
\usepackage[hidelinks]{hyperref}

\usepackage{balance} 
\usepackage{algorithm}
\usepackage{algorithmic}

\title{Justifying Social-Choice Mechanism Outcome for Improving Participant Satisfaction}

\author{Sharadhi Alape Suryanarayana \thanks{sharadhi.as@gmail.com}}

\author{David Sarne \thanks{david.sarne@biu.ac.il}}

\author{Sarit Kraus \thanks{sarit@cs.biu.ac.il}}

\affil{Bar-Ilan University, Israel}

\begin{document}
\date{}
\maketitle

\begin{abstract}
In many social-choice mechanisms the resulting choice is not the most preferred one for some of the participants, thus the need for methods to justify the choice made in a way that improves the acceptance and satisfaction of said participants. One natural method for providing such explanations is to ask people to provide them, e.g., through crowdsourcing, and choosing the most convincing arguments among those received. In this paper we propose the use of an alternative approach, one that automatically generates explanations based on desirable mechanism features found in theoretical mechanism design literature.  We test the effectiveness of both of the methods through a series of extensive experiments conducted with over 600 participants in ranked voting, a classic social choice mechanism. The analysis of the results reveals that explanations indeed affect both average satisfaction from and acceptance of the outcome in such settings. In particular, explanations are shown to have a positive effect on satisfaction and acceptance when the outcome (the winning candidate in our case) is the least desirable choice for the participant. A comparative analysis reveals that the automatically generated explanations result in similar levels of satisfaction from and acceptance of an outcome as with the more costly alternative of crowdsourced explanations, hence eliminating the need to keep humans in the loop. Furthermore, the automatically generated explanations significantly reduce participants' belief that a different winner should have been elected compared to crowdsourced explanations.
\end{abstract}





         
\newcommand{\BibTeX}{\rm B\kern-.05em{\sc i\kern-.025em b}\kern-.08em\TeX}







\section{Introduction}

As intelligent agents and AI-based systems are becoming increasingly prevalent and with the increase in their use, the ability to explain their decisions and choices to their human users becomes a standard requirement. The need for such an explanation is especially acute whenever 
these systems employ complex algorithms that are generally incomprehensible to the non-expert users, or whenever it is difficult for users to distinguish bad outcomes from good decisions (given the information available to the system). For example, when providing incorrect GPS directions \cite{wolfe2014driving}, task division that is not preferred by the user \cite{zahedi2020not} or actions picked based on incorrect modelling assumptions leading to a catastrophe (as in the case of the mortgage crisis \cite{donnelly2010devil}). 
In all of these examples, convincing the stakeholders that the decision is justified needs to consider two factors - the complexity of the algorithm and the dissatisfaction of the stakeholders - thus making the said process difficult.
Indeed, in recent years researchers have invested tremendous efforts in developing methods for generating explanations 
for decisions made by machine-learning-based systems  \cite{lucic2020does, ray2019can} and recommender systems \cite{kleinerman2018providing, kouki2019personalized}, which are mainly black-box algorithms. Common to all of these methods, is that they are ultimately trying to convince the user of the optimality of the decision made according to some well-defined goal function which is known to and accepted by the user.

In many settings, however, the system is designed to reach a decision that applies to several participants, often associated with conflicting goals. This is the case, for example, in social choice settings (e.g., fair division, auctions, voting systems), for which numerous mechanisms have been designed over the years. Here, explaining the decision made by the system is far more complicated, as it is clearly not optimal for some of the participants. The purpose of the explanation is thus to increase participant satisfaction from and acceptance of the choice made whenever the decision made is not in her favor. As such, explanations should be framed within the social context of the underlying setting, emphasizing aspects such as fairness, pareto optimality and overall social welfare \cite{kraus2020ai}. Despite the importance of providing explanations in such settings, most work to date, as we review later on, has focused on explaining decisions of the above first type, i.e., arguing the optimality and correctness of the decision made in the context of the user's own goals. Methods for arguing the legitimacy of selections made in social choice settings are quite limited.

This paper studies the use of explanations for justifying the outcome of such social choice mechanisms. In particular, we devise and extensively evaluate two methods for generating explanations. The first uses crowdsourcing, leveraging human intelligence in order to produce explanations that are likely to convince people of the legitimacy of the winning choice. This follows the line of work where human intelligence is used to improve the performance of system explainability \cite{kulesza2015principles, smith2020no, zar2021explaining}. The second method automatically generates explanations based on features considered desirable when designing a mechanism for making the social choice. For this, we make use of the heterogeneity of different mechanisms, axioms and criteria suggested in literature. The explanations generated are centered around several numerical features that are used to quantify the aforementioned theoretical concepts, thus eliminating the need for the explanations to focus on a single argument. This latter method, if successful, can save the extensive human resources required in the first (and more intuitive) approach. 

We use ranked voting as an infrastructure for evaluating the two methods. Voting rules are used to decide diverse yet commonplace subjects such as meeting schedules, award recipients, holiday destinations and representatives of countries. These settings are known as \emph{Preference Aggregation Settings}. Despite their widespread usage, voting rules suffer from a couple of drawbacks. The first is that in all cases where individuals rank their preferences, it is impossible to determine a clear order of preferences while adhering to mandatory principles of fair voting procedures \cite{arrow1951social}. This is unlike several other decision-making settings where a solution can be reached while optimizing a certain measure of social welfare.
Consequently, there are multiple voting rules being used today, which leads to the second drawback - the fact that different voting rules may lead to different winners and the results may not be fully understood or accepted by the voters. Our experiments use six varied instances of ranked voting, differing in the vote distribution and the winner picked. Since the goal is to test for the effectiveness of the explanations, in each such instance, the participants are asked to rate their satisfaction and acceptance of the outcome that is not their most preferred, with and without the generated explanations.

The analysis of the results of the experiments conducted with 465 participants reveals that explanations can be highly effective in improving participant satisfaction from and acceptance of results in social settings.   Furthermore, the performance of the feature-based, automatically generated explanations, i.e., those that do not require manual labor to extract, is as good as the performance of explanations suggested by people through crowdsourcing. In particular, the feature-based explanations significantly reduce the percentage of participants who believe another candidate is more deserving of winning, compared to crowdsourced explanations and no explanations. 

\section{Explanation Generation}
We consider two explanation generation methods. The first relies on the wisdom of the crowd, whereas the second attempts to automatically generate explanations based on properties and features of such mechanisms. Both methods use two primary steps: generating the explanations and then evaluating them based on some scoring or ranking process. The output of each method is up to $n$ explanations.

\subsection{Crowdsourced Explanations}

Even though the word ``crowdsourced" is self-explanatory as to how the explanations are obtained, we propose a refined approach for obtaining the best set of explanations while relying on human intelligence. 

\noindent\paragraph{Explanation Generation} Participants are presented with the relevant social choice setting, including the winning choice. After making sure they fully understand and internalize the setting, they are asked to provide some pre-defined number of reasons in support of the winning choice.

\noindent\paragraph{Explanation Ranking} A different set of participants are presented with the relevant social choice setting, including the winning choice. After making sure they fully understand and internalize the setting, they are asked to review the explanations received in the former stage, and pick the $w$ most convincing explanations. The explanations are then ordered according to the number of times they were picked by the reviewing participants and the set of $n$ top explanations in the ordered list is the output of the process.

\subsection{Feature-based Explanations}

For the feature-based explanations we use two similar stages, though their content is completely different. 

\noindent\paragraph{Explanation Extraction} A natural source for explanation of a mechanism outcome is the set of properties that prior work on mechanism design aimed at optimizing or satisfying. The intuition is simple, suggesting that the properties of a mechanism, even if they hold only with purely rational and non-computational bounded agents, are likely to be important, to some extent, to people. For example, when considering a solution produced by a cake cutting mechanism, desirable properties of the solution include fairness -- envy-freeness, equitability and proportionality \cite{brams2012maxsum}, efficiency -- non-wastefulness, Pareto efficiency and utilitarian-maximality \cite{ianovski2012cake}. When considering a solution produced by a resource allocation mechanism, desirable properties will include, among others, budget-balancing, envy-freeness,  efficiency of the allocation and minimal manipulability \cite{procaccia2018fair, andersson2014budget}. Each such property can be used as a potential explanation for the selection of a solution that satisfies it using a feature that quantifies it. Obviously, some of the possible solutions of a mechanism may fall short in comparison to at least one other solution with respect to every feature.
While justifying the former set of solutions is an intriguing problem, it is beyond the scope of the current paper, as in real-life it is very unlikely that such an inferior solution will be picked to begin with.  
For solutions satisfying $k\leq n$, where $n$ is the number of required explanations and $k$ is the number of features considered as potential explanations, the natural approach would be to present the corresponding $k$ explanations. For solutions satisfying $k>n$, we use a heuristic scoring function (see the following paragraph) that helps in determining which $n$ of the $k$ explanations will be presented (and in what order). 

\noindent\paragraph{Explanation Scoring} Our heuristic explanation scoring aims to measure how apparent the dominance of a given solution is when measured with respect to a given desired property. For example, consider the budget-balancing property of a solution to an allocation problem. Here we can measure how far each solution is from budget balancing (with a zero-distance in case of fully satisfying the requirement).\footnote{Meaning that even if the feature is binary, we can measure how close we are to having it hold. For example, take the envy-free criterion. Two solutions can be envy-free, and yet we can allegedly measure the extent of holding it  (e.g., measure how far we are from ``envy" by measuring to what extent  the allocation should change so that one of the participants envies another).}  Consider the set of solutions $S=\{S_1,...,S_w\}$. For each solution $S_i$ we use $d^j_i$ to denote the extent to which it satisfies feature (based on the property) $j$. We measure the dominance of $S_i$ over the other solutions, with respect to feature $j$, as $\sum_{t=1,t\neq i}^w |d^j_i-d^j_t|$.  This can be interpreted as how far ahead the solution is with respect to feature $j$, compared to the other solutions. We emphasize that various alternatives may be considered here. For example, one may choose to measure the difference compared to the second best solution with respect to the feature considered, to avoid biases resulting from the existence of other extremely bad solutions and such.  Still, we find the sum of differences to be a decent measure as it relates to all other solutions. Since different features can be measured over different scales, we normalize their score by dividing the dominance measure by the maximum (or minimum, depending on the nature of the feature) potential value a solution may obtain.\footnote{For example, when considering the first place votes (plurality) criterion, the maximum is the number of total voters.}  Finally, the explanations are ordered according to their normalized score and the top $n$ explanations (with ties broken randomly) in the ordered list are the output of the process. The process is summarized in Algorithm \ref{alg:cap}.

\begin{algorithm}\label{alg}
\caption{Picking Feature-Based Explanations}\label{alg:cap}
\begin{algorithmic}[1]
\REQUIRE Set of solutions $S$, Solution to explain $S_i$, number of explanations required $n$, Set of features $J$
\STATE $temp \gets \emptyset$
\FORALL {$j \in  J $} 

\FORALL{$S_t \neq S_i \in S$} 
\STATE Obtain the value of $d^j_i$
\STATE Obtain the value of $d^j_t$
\IF{$d^j_i$ is inferior to $d^j_t$}
\STATE $J \gets J-j$
\STATE \textbf{break}
\ENDIF
\ENDFOR
\ENDFOR

\FORALL{$j \in  J $}
\STATE Calculate $score^j_i = \sum_{t\neq i} |d^j_i-d^j_t|$
\STATE Normalize $score^j_i$ (divide by max/min potential score)
\STATE $temp = temp \cup (j, score^j_i)$
\ENDFOR
\STATE \textbf{sort} $temp$ by the values $score^j_i$
\STATE\textbf{Return} $top-n$ features of $temp$
\end{algorithmic}
\end{algorithm}

\noindent\paragraph{Comparison of Methods} Obviously the crowdsourced-based  method is the more intuitive and natural one, as explanations proposed by people are likely to be highly appealing to the participant. Still, this method requires substantial resources in the form of payment to those proposing the explanations. These resources should be spent for every new instance of the problem. The feature-based approach, on the other hand, requires only the extraction of the features themselves (e.g., based on the guarantees of the different mechanisms proposed for making the decision). This is done only once and can then be used in generating explanations for every new instance.

\section{Ranked voting}\label{sec:Voting}

We test the explanations produced with the two above methods for different winners in ranked voting. Ranked voting uses preference ballots in which the voter ranks the choices in order of preference. It is used as a means of deciding upon the winner of an election. The most important differences between ranked voting systems lie in the methods used to decide which candidate (or candidates) are elected from a given set of ballots.

Formally, denote $V = \{ v_1, v_2,...v_N \}$ as the set of $N$ voters and $C = \{c_1, c_2, ..., c_m \}$ as the set of $m$ candidates. Let $Q_i(x) \in \{1,2,..m\}$ be the rank of candidate $x \in C$ with respect to voter $v_i$. We say that voter $v_i$ prefers candidate $x$ to $y$ is denoted $x \succ y$, if $Q_i(x) < Q_i(y)$. In order to conceptualize and quantify the notion of preference, we use the concept of cardinal utilities as a function $U_i: C \to \mathbb{Q}$. $U_i$ is consistent with the preference of voter $v_i$ such that $U_i(x) > U_i(y)$ if and only if $x \succ y$. The usage of cardinal utilities translates as: if voter $v_i$ prefers candidate $x$ to candidate $y$, then the utility of $v_i$ from $x$ is higher than that of $y$. Let $L$, also called preference ballot, denote the preference orders of all of the $n$ voters. A voting rule is a function $f:L \to C$ that returns the set of winning candidates given the voting profile. 

Voting rules such as Plurality, Plurality with Runoff, Borda and Condorcet are a few of the many voting rules used today in ranked voting. They all have some desirable theoretical properties, but none of them is perfect. Procaccia \cite{procaccia2019axioms} argues that these desirable properties, many of which are expressed as axioms, can be used to explain the results of any mechanism and not just to cater to the needs of the designer.  

In order to design explanations to justify the selection of the winning candidate in this domain, we make use of the following features:

 \begin{enumerate}
     \item \emph{First Place Votes (Plurality)}: The total number of first place votes that the candidate received. This feature is mainly used in voting methods such as Plurality and Majority \cite{hazon2012evaluation}.
     \item \emph{Bucklin Score}: The total number of votes for the first place for each of the candidates is counted. If no candidate gets more than fifty percent of the votes, the cumulative number of votes for the first and second place is counted. If any candidate secures more than half of the votes, the counting stops, otherwise it goes on at every level of preference until at least one candidate gets more than half of the total votes. The candidate that gets the highest cumulative score is declared the winner and the cumulative score is called the Bucklin Score \cite{hazon2012evaluation}.
     \item \emph{Borda Count}: A candidate's votes at every position are weighted in reverse proportion to the ranking. The sum of the weighted score is the Borda Count \cite{emerson2013original}.
     \item \emph{Head-to-head Comparison}: For each pair of candidates, the candidate which is preferred by the highest number of voters is determined. The candidate that wins every such comparison is the winner \cite{young1988condorcet}. 
     \item \emph{Last Place Votes}: The total number of votes cast ranking the given candidate in the last place. This feature is obtained from the criterion called ``Absolute Loser Criterion", which states that if a candidate has the highest last place votes, the candidate should not win \cite{mccune2019can}.
      \item \emph{Greatest Pairwise Opposition}: The greatest score against a given candidate when compared pairwise with each of the other candidates. This is used in the Minimax Condorcet Method where the winner has the least greatest pairwise opposition \cite{felsenthal2018voting}. 
\end{enumerate}

\section{Experimental Framework}

\begin{figure}[ht]
	\centering
	\includegraphics[width=1\linewidth]{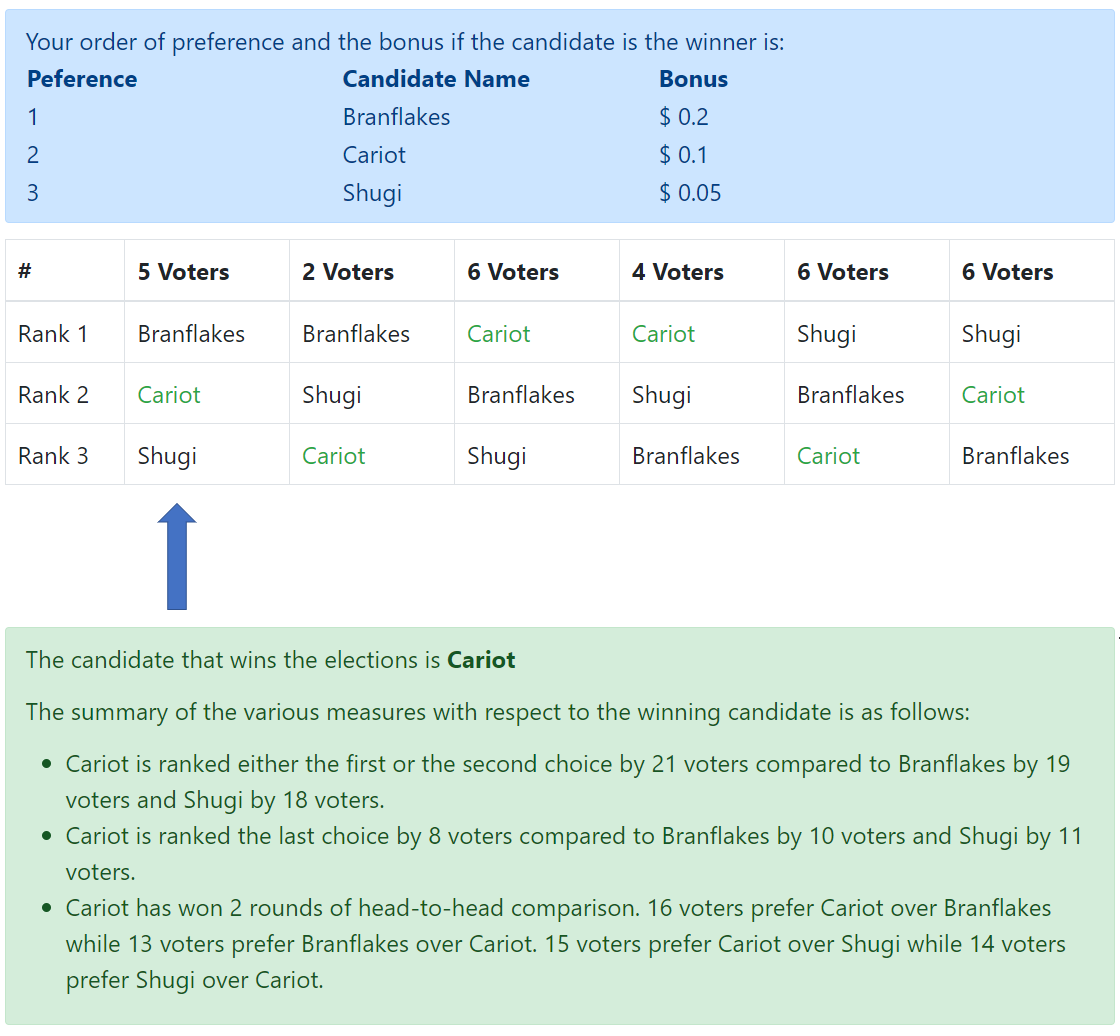}

	\caption{Screenshot: example of the voting table, winning candidate and explanations.}

	\label{fig:explanationswithvotingtable}	
\end{figure}

As a framework for our experiments, we used an interactive web-based application that emulates a ranked voting system.  The system presents to each participant her specific preference order of candidates and assigns a monetary compensation to each, accordingly. It then presents the preferences of other voters, in a tabular form, aggregating those with similar preference orders (see screenshot in Figure  \ref{fig:explanationswithvotingtable}). 
The participant's preference aligns with any of the 4 columns where the first choice is not the winning candidate, i.e., she takes the role of one of the voters voting for that order. Once the participant finishes carefully checking the ranked voting instance, the system announces the winning candidate (without disclosing the winner determination rule). This stage can be accompanied with the presentation of an itemized list of explanations. Finally, the participant is asked to specify her satisfaction and acceptance of the result (on a scale of 1-5). The participant, in case of feeling the winning candidate is not the right one, can choose an alternate winner.

\section{Experimental Design}

\noindent \paragraph{Voting domain} In our experiments participants were introduced with the results of a ranked voting instance aiming to determine the best out of three cereal brands, namely Branflakes, Cariot and Shugi, that are used in a distant country. This was done in order to mitigate the effect of any personal bias, forcing participants to rely solely on the preference order provided for them.

\noindent \paragraph{Voting instances}
Each voting instance was comprised of $m=3$ candidates and $N=29$ voters, where the participant is one of them. This enabled sufficient variations of voting tables and helped avoiding round numbers (numbers ending with $0$ or $5$) of the total number of voters in the poll, as such numbers may be treated differently by participants \cite{mitchell2001clustering}. Having three candidates, we had a total of six different preference permutations across which to divide the 29 votes. 

Six voting instances were manually generated, each specifying a certain division of votes and the winner picked. These are described in Table  \ref{tab:instances} which also specifies the features that hold for the winner in each instance according to the list provided in Section \ref{sec:Voting}.\footnote{Note that overall we had a total of four different vote distributions where two of them were used twice, each time with a different winner.} The vote distribution column specifies the number of votes assigned to each of the permutations according to their order as given in Figure \ref{fig:explanationswithvotingtable}. 
These instances capture diverse settings like the winning candidate standing first with respect to all of the six features (Voting instances 1 and 2), the winning candidate standing first with respect to only one feature (Voting instances 3 and 4a) and the winning candidate not having the highest first place votes (Voting instances 3a and 4). This diversity gives rise to situations where the number of crowdsourced explanations is higher than the number of feature-based explanations (Voting instances 3 and 4a), giving an advantage to crowdsourced explanations, and those where the number of feature-based explanations and crowdsourced explanations are equal (Voting instances 1, 2, 3a and 4).

\begin{table}[ht]
\caption{Vote distribution and the different features satisfied by the winner picked for that instance: (P)lurality, (B)orda, (H)ead-to-Head, (BR) Bucklin Rule , (M)inimax Condorcet, (L)ast.} 
\label{tab:instances}
\begin{tabular}{|c|c|c|}
\hline
Instance & Vote Distribution  & Winner Features \\ \hline
1        & 6/4/4/7/4/4 & P/B/H/BR/M/L    \\ \hline
2        & 6/2/8/5/4/4 & P/B/H/BR/M/L    \\ \hline
3        & 1/4/7/6/7/4 & P               \\ \hline
3a       & 1/4/7/6/7/4 & B/H/BR/M/L      \\ \hline
4        & 5/2/6/4/6/6 & B/H/BR/M/L      \\ \hline
4a       & 5/2/6/4/6/6 & P               \\ \hline
\end{tabular}
\end{table}

\noindent \paragraph{Participant Preferences}
A mere presentation of the voting table and the winning candidate with no selfish interest might have resulted in a lukewarm response from the participants. In order to interest them in the voting table and the winning candidate, the concept of utility from the winning candidate had to be realized. Consequently, the participants were allotted a preferred order of candidates. The participants were told that they would be awarded a monetary compensation (bonus, on top of the base payment) based on the ordinal position of the winning candidate with respect to the aforementioned order of preference.
However, for the possibility of feeling compelled to be satisfied if the most preferred candidate of the participant won (consequently, getting the highest bonus), i.e. positive reciprocity bias \cite{fehr2000fairness}, the participant was allotted an order of preference such that the winning candidate is never the first preference. 
Hence, the winning candidate in the experiments was either the participant’s second preference or third preference. 
The bonus was the least if the winning candidate were the participant's third preference. In comparison to the least bonus, the bonus was doubled if the participant's second preference was declared winner and quadrupled if her first preference was picked.
Since the ordinal position of the winning candidate with respect to the participant's order of preference is a factor in our analysis, we will henceforth refer to it as \emph{Preference Index}.

\noindent \paragraph{Experimental Flow}
Participants were given a brief introduction of the task which was followed by obtaining their Informed Consent and the Demographic Details. The participants were then given detailed instructions on the different phases of the task including the voting table, the allocation of the preferred order of candidates, the correlated bonus and the final survey. They were also informed that the survey had to be answered assuming that the participant was one of the voters. Prior to moving on to the main task, in order to acquaint the participants with the different concepts, a sample voting table was displayed and a quiz based on the same was presented. The participants were then shown the actual voting table.

\noindent \paragraph{Measures}
Based on the voting table and the winning candidate, the participant was required to answer three questions relating to her acceptance of and satisfaction from the result: 
\begin{itemize}
    \item \textbf{Satisfaction}. (Question 1: How satisfied are you with the SELECTION MECHANISM ending up with <winner identity> as the winner, given the data presented to you?)
    \item \textbf{Acceptance}. (Question 2: Do you think <winner identity> was rightfully elected?)
    \item \textbf{Alternate Choice}. (Question 2a: I believe option <select> should have been elected, based on the data). This question was asked only if the participant responded to Question 2 above as ``No" or ``Not Necessarily" (values 1 and 2 on the scale) and the options that became available to the participant were the two non-winning candidates.
    \item\textbf{Disappointment}. (Question 3: How do you feel that your first preference, <participant's first preference>, has not won and you have not gotten the maximum bonus?)
\end{itemize}
All three questions are answered by picking a value on a Likert scale of 1-5. The first two questions were used for the analysis on the effect of explanations on the participant with 1 being the least desired value (e.g. lowest level of satisfaction) and 5 being the most desired value. The third question was used primarily for a sanity check (1 - Very unsatisfied and 5 - Highly Satisfied) -- anyone answering it as ``highly satisfied" or ``satisfied" (values 4 and 5 on the scale) had her responses purged. 

\noindent \paragraph{Participants} Participants were recruited and interacted through Amazon Mechanical Turk (AMT). Participation was restricted to AMT workers from the United States who had completed more than 1000 Human-Intelligence-Tasks (HITs) and had more than $98$ percent of their participated HITs approved. IRB Approval was obtained in order to conduct the experiment. 
To prevent any carryover effect, each participant was assigned to one voting table instance and a corresponding winning candidate. 

\paragraph{Explanations} 
The number of explanations to be presented was limited to three in order to avoid an information overload \cite{pu2007trust}.\footnote{While \cite{pu2007trust} is in the field of recommender systems, we wanted to test a similar approach in the field of social choice theory.} The features used for providing the explanations were those detailed in Section \ref{sec:Voting}. 

For crowdsourced explanations we used AMT workers. Each of the six instances was presented to 15 different workers according to the above procedure, except for assigning them with an order of preference. After carefully going over the voting table, each worker had to provide three reasons to support the selection of the predetermined winning candidate.  
The crowdsourced explanations collected were then presented to other AMT workers (20 workers for each of the six instances). These workers followed the same procedure of becoming acquainted with the problem instance, except that their task was to pick the three most convincing explanations in the context of justifying the selection of the pre-picked winner. This procedure resulted in a subset of three explanations that were picked the most by the 20 workers for each problem instance. 

Table \ref{table:ExplanationClass} details the type of explanations generated with the two methods tested. Here, the symbol ``N" in the crowdsourced explanations column stands for an explanation that is not related to any of the specified features (e.g., ``The biggest group and second biggest group of voters vote for X as their first choice"). We note that the extent of overlap between the explanations generated with the two methods highly varies - for some instances the explanations are quite similar (e.g., instances 3a and 4) while for others they are completely different (e.g., instance 2). Instance 2 has four possible feature-based explanations due to a tie in the scores of ``L" and ``BR" which is broken randomly. For some of the instances (e.g., instances 3a and 4a) the crowdsourced explanations rely on features for which the selected outcome is not even the best among the three. Finally, we note that with the crowdsourced explanation we find some redundancy (providing the same explanations in different words), e.g., in instances 1 and 3, two of the explanations given relate to the advantage of the winning choice in terms of first preference votes (plurality). 

\begin{table}[ht]
\caption{Classification of the crowdsourced-based and feature-based explanations for the different instances.}
\label{table:ExplanationClass}
\begin{tabular}{|c|c|c|c|}
\hline
Instance & Winner Features & \multicolumn{1}{c|}{Crowdsourced } & \multicolumn{1}{c|}{Feature-based} \\ \hline
1        & P/B/H/BR/M/L    & P/P/BR                           & H/L/BR                             \\ \hline
2        & P/B/H/BR/M/L    & P/N/N                            & H/M/L/BR                               \\ \hline
3        & P               & P/P/N                            & P                                  \\ \hline
3a       & B/H/BR/M/L      & BR/H/P                           & H/BR/L                                    \\ \hline
4        & B/H/BR/M/L      & BR/H/N                           & L/BR/H                                   \\ \hline
4a       & P               & P/BR/H                           & P                                    \\ \hline
\end{tabular}
\end{table}

\paragraph{Study Design}
The study used a $3 \times 2$ between-participants experimental design, with factors of \emph{Explanation} -- None, Crowdsourced and Feature-based -- and \emph{Preference Index} -- Second Preference and Third Preference. While there are six diverse problem instances, we do not include them as a factor in the Analysis of of Variance (ANOVA) as the aim of presenting the participant with explanations is to overcome this diversity. The analysis also does not compare satisfaction and acceptance between different problem instances and whenever applicable, different methods within a specific problem instance are compared. There are approximately 80 responses per \emph{Preference Index} and \emph{Explanation}.

\paragraph{Data and Analysis}
Overall, we had 465 participants in our experiments according to the breakdown given in Table \ref{table:participantbreakdown}, which also includes participants' demographics.
The analysis of participants' average satisfaction and acceptance was carried out using $3 \times 2$ ($Explanations \times Preference Index$) ANOVA with Aligned Rank Transforms, since such a test is more appropriate in a setting where Likert scale data is analyzed \cite{10.1145/1978942.1978963}. 
Further tests on the efficacy of the explanations were carried out by splitting the data on the basis of the  \emph{Preference Index}, using an upper-tailed Mann-Whitney-Wilcoxon test. The comparison between the feature-based explanations and crowdsourced-explanations was carried out using a two-sided Mann-Whitney-Wilcoxon test. Finally, to analyze the difference between the proportion of participants convinced that no other candidate could be justifiably selected (i.e., based on the answers to question 2a), a 2-sample Z-test for proportions was carried out.
\begin{table}[ht]
\caption{Breakdown of participants by explanation, average age and division (in percentages) between men and women.}
\label{table:participantbreakdown}
\begin{tabular}{|c|l|l|l|}
\hline
Explanation         & \#Participants  & Average Age &\%Men/\%Women            \\ \hline 
None     &      158             &39.10          &58.2/41.8 \\ \hline
Feature-based  &      154            &41.68          &58.3/41.7    \\ \hline
Crowdsourced   &      153            &42.16          &50.3/49.7    \\ \hline
\end{tabular}
\end{table}

\section{Results and Analysis}
\begin{figure}[ht]
	\centering
	\includegraphics[width=0.8\linewidth]{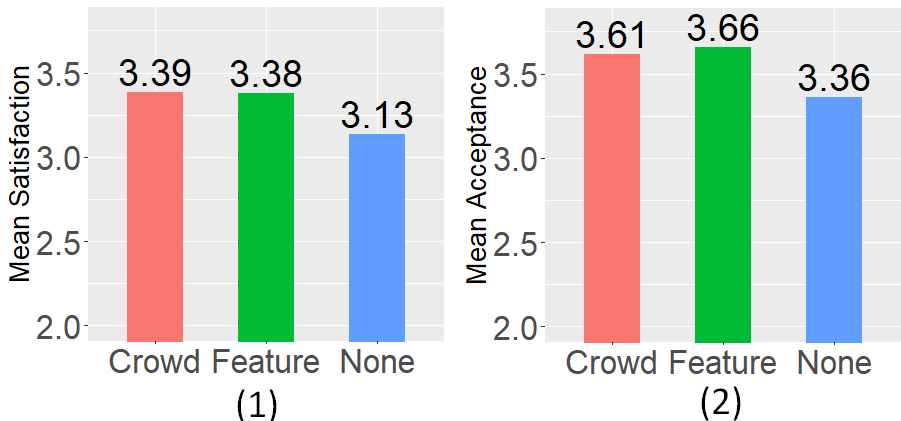}
   
	\caption{Average Satisfaction (1) and Acceptance (2) across the three types of Explanations.}
	\label{fig:OverallSatAndAgreement}	
\end{figure}

Figure \ref{fig:OverallSatAndAgreement} depicts the average satisfaction and average acceptance cross-instances when not providing explanations and when providing crowdsourced and feature-based explanations. From the figure we observe that both types of explanations provided resulted in a substantial increase in satisfaction and acceptance compared to not providing explanations. We emphasize that even though the explanations provided managed to eliminate $13-22\%$ of the difference between the scores obtained with no explanations and a perfect satisfaction/acceptance score of 5, it is very unlikely that even a perfect explanation could attain that perfect score, given that it is the participant's second and third preference that is picked as the winning outcome.

\begin{figure}[ht]
	\centering
	\includegraphics[width=1\linewidth]{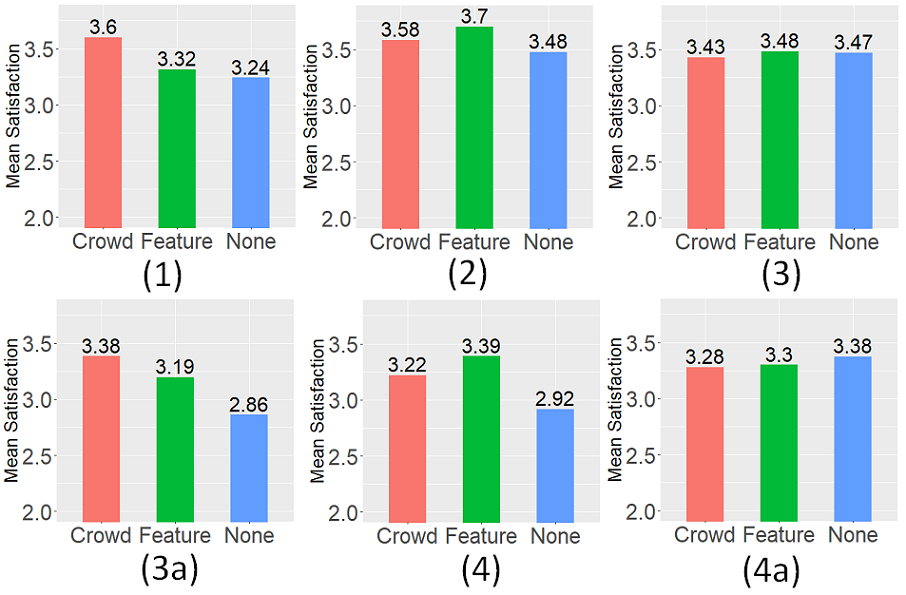}
   
	\caption{Average Satisfaction across the 6 instances.}

	\label{fig:All6Satisfaction}	
\end{figure}

\begin{figure}[ht]
 	\centering
 	\includegraphics[width=1\linewidth]{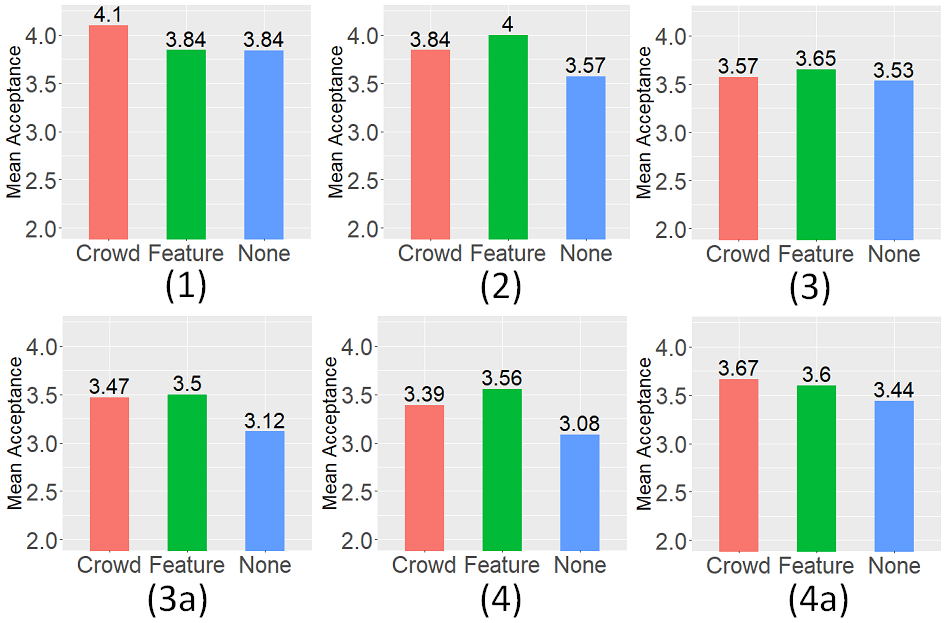}
    
 	\caption{Average Acceptance across the 6 instances.}
 	\label{fig:All6Acceptance}	
 \end{figure}

From the ANOVA (taking into consideration the effect of the Preference Index as well as the interaction effect of the Preference Index and the explanations) we obtain that the increase in average satisfaction and acceptance resulting from the provision of explanations, with both explanation-generation methods tested, is statistically significant (p-value<0.05). To better understand where explanations make the most difference we provide the average satisfaction (Figure \ref{fig:All6Satisfaction}) and acceptance (Figure \ref{fig:All6Acceptance}) for each of the 6 instances in Table \ref{tab:instances}.  From the figures we observe that for instances 3 and 4a, where the winning candidate has the highest first place votes (i.e., the plurality criterion holds), the presence of explanations does not improve a participant's satisfaction and only slightly improves acceptance.  Indeed, the Plurality voting rule enjoys higher familiarity than the others among people \cite{mccune2019can} and was found to be dominant in experiments studying people's vote manipulation and bias towards certain candidates \cite{bassi2015voting, forsythe1996experimental,tal2015study}. When the winning candidate does not have the highest first place votes, the choice needs to be explicitly justified for people to be satisfied and accepting.  Explanations did improve in all of the other four instances. In fact, the highest improvements compared to providing no explanations were obtained with instances 3a and 4, where plurality is not a feature to begin with. These insights suggest that the scoring and ranking of explanations can be improved by taking into consideration participant familiarity with the different features, but this is beyond the scope of the current paper.

Another important insight revealed from Figures \ref{fig:All6Satisfaction} and  \ref{fig:All6Acceptance} is that participants' satisfaction from and acceptance of the winning candidate in a given instance are not fully correlated.  For example, in instance 4a, having no explanations resulted in a slightly elevated satisfaction, however a reduced acceptance compared to when provided with explanations. Similarly, with explanations there are instances (e.g., 3a) where feature-based explanations improved acceptance (compared to crowdsourced explanations) however decreased satisfaction, and vice-versa (e.g., 4a).  This suggests that the choice of the explanations to be used should depend on the system designer's goal of whether to increase satisfaction or acceptance.

\begin{figure}[ht]
	\centering
	\includegraphics[width=0.8\linewidth]{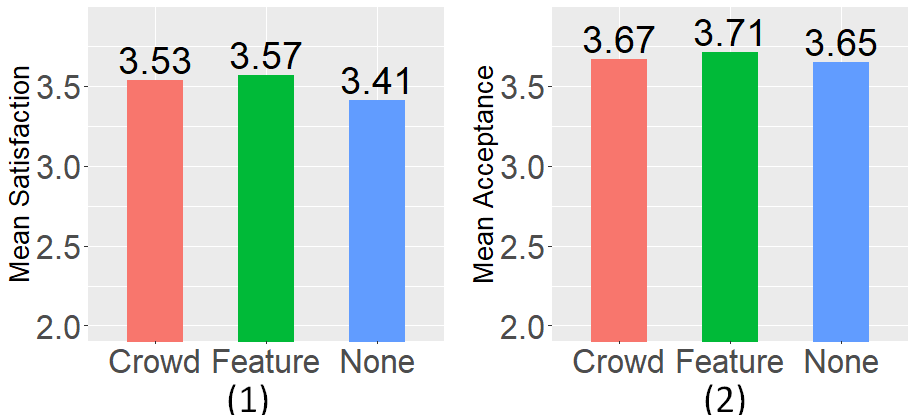}
    
	\caption{Average satisfaction (1) and acceptance (2) across the three types of Explanations when the winning candidate is the second preference of the participant.}
	\label{fig:OverallSatAndAgreementSec}
\end{figure}

\begin{figure}[ht]
	\centering
	\includegraphics[width=0.8\linewidth]{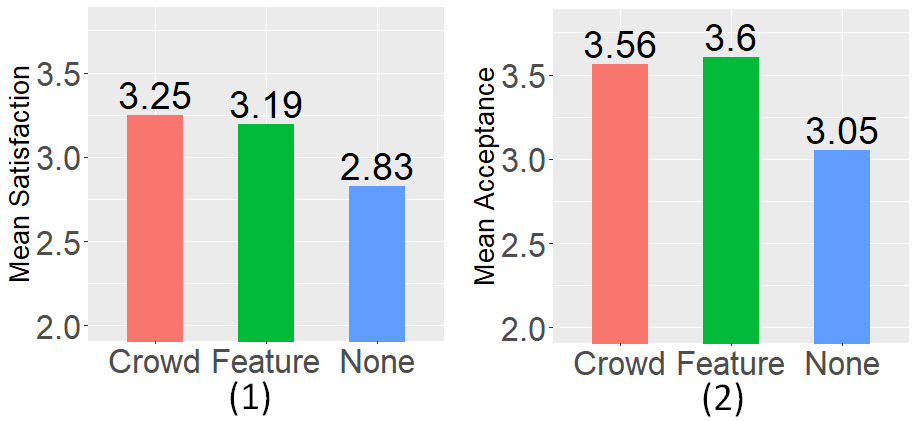}
  
    	\caption{Average satisfaction (1) and acceptance (2) across the three types of Explanations when the winning candidate is the third preference of the participant.}
	\label{fig:OverallSatAndAgreementTh}	
\end{figure}

To better understand the effect of the participant's preference index of the winning candidate on satisfaction and acceptance in the context of providing explanations, we introduce Figures \ref{fig:OverallSatAndAgreementSec} and \ref{fig:OverallSatAndAgreementTh}. These figures are equivalent to Figure \ref{fig:OverallSatAndAgreement}, except that they distinguish only participants who had their second choice declared winner (Figure \ref{fig:OverallSatAndAgreementSec}) and those that had their third choice declared winner (Figure \ref{fig:OverallSatAndAgreementTh}). From the figures we observe that satisfaction and acceptance are both affected by the ordinal position of the winning candidate in the participant's allotted order of preference, i.e., the Preference Index---
the perceived increase in satisfaction and acceptance is significantly greater when the winning candidate is the participant's third preference, compared to when it is the second preference. Indeed, when the participant has the least utility from the winning candidate , i.e., when she is likely to be most disappointed, explanations are an effective tool in improving the satisfaction and acceptance.

\begin{figure}[ht]
	\centering
	\includegraphics[width=0.9\linewidth]{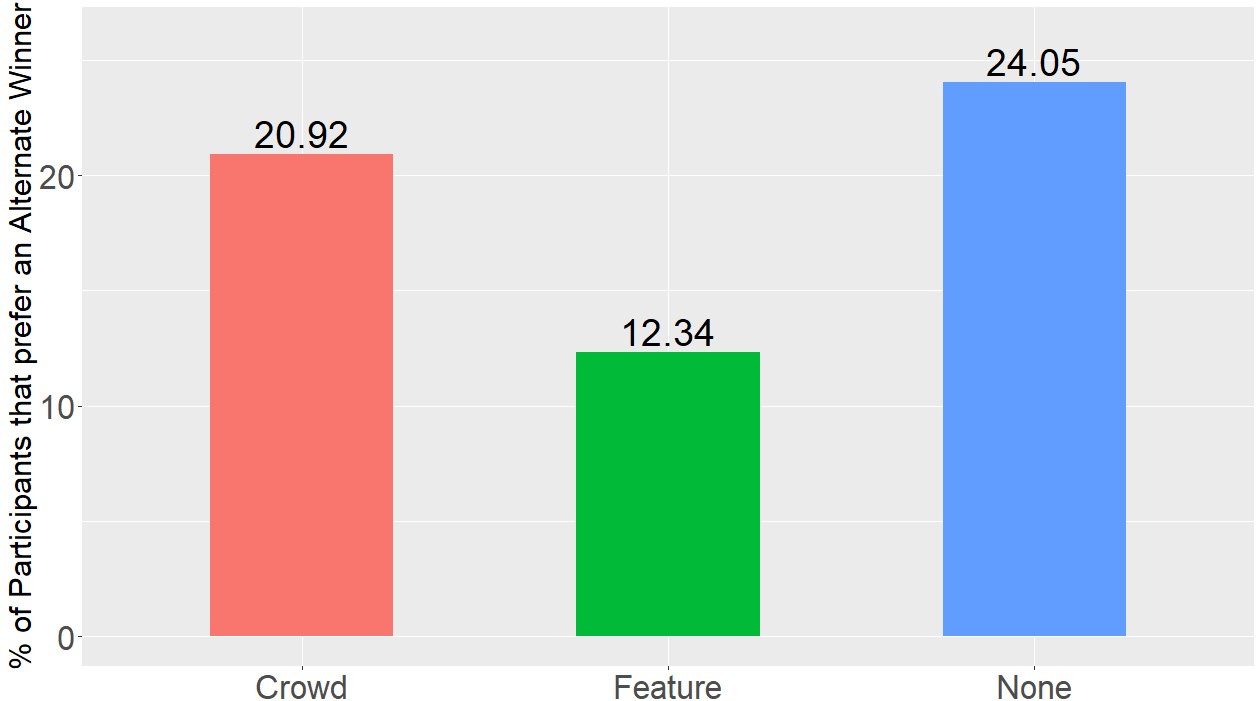}

	\caption{Percentage of participants that feel that another winner is justified.}

	\label{fig:PercentageDisappointed}
\end{figure}

A pairwise comparison of the two explanation-generation methods (Figures \ref{fig:All6Satisfaction} and \ref{fig:All6Acceptance}) reveals that the feature-based explanations manage to perform at least as well as crowdsourced explanations. However, we observe that the performance of the two explanation methods highly varies between instances (except for instances 3 and 4a). This means that the choice of the explanation to be provided matters, and it is not simply the provision of explanations itself that accounts for the improvement achieved in acceptance and satisfaction. Figure \ref{fig:PercentageDisappointed}  depicts the percentage of participants who believed that one of the other candidates was more deserving of being the winner, for the case where no explanations are provided and with the two explanation-generation methods tested. From the graph we observe that providing explanations decreases the percentage of participants who believe a different winner should have been elected. Furthermore, while the decrease obtained with the crowdsourced explanations is not statistically significant (p-value>0.05), with the feature-based explanations the improvement is statistically significant (p-value< 0.05), both compared to when providing crowdsourced explanations and when providing no explanations.

\section{Related Work}

Our work falls primarily under the category of explaining the decisions of intelligent systems.  This line of work has gained a lot of traction in recent years due to the European Union's new General Data Protection Regulation (GDPR), which highlights the right of citizens to receive meaningful information on the algorithmic decisions involving them \cite{goodman2017european,selbst2018meaningful}. 
Since black-box models (such as deep learning) have been a cause of caution \cite{10.2307/j.ctt13x0hch} and have led to the demand of transparency in algorithmic decision-making \cite{doshi2017accountability}, the more-researched domain in the context of explainability has been Machine Learning. This has resulted in a distinct sub-field called Explainable Artificial Intelligence (XAI).  With the advent of this field, work on defining the concepts \cite{rosenfeld2019explainability}, establishing a framework to assess explainability systems \cite{sokol2020explainability, hall2019systematic, verhagen2021two} and identifying the best practices for transparency in algorithmic systems \cite{schelenz2020best} have been carried out.  Ideas from other fields such as psychology \cite{miller2019explanation}, multi-agent systems \cite{ciatto2020agent}, argumentation \cite{amgoud2009using}, planning \cite{sohrabi2011preferred} and logic \cite{calegari2020explainable} have also been used to establish definitions for explainability and propose methods to construct explanations.

With respect to the specific case of social choice theory, work on explaining and justifying collective decisions is far more limited. For example, some work suggested explanations 
using axioms \cite{cailloux2016arguing,boixel2020automated}, justifying the decisions of approval sorting \cite{belahcene2018accountable} and analysis of the computational complexity of generating explanations \cite{boixel2021complexity, NEURIPS2020_10c72a9d}. However, translating the techniques of these studies to conducting real-time experiments with non-expert humans is challenging, due to its theoretical nature.

The necessity to translate theoretical foundations into explanations was put forward by Procaccia \cite{procaccia2019axioms}. This is justified by an experiment with actual stakeholders of a rent division problem to whom the solutions are explained using axiomatic foundations \cite{gal2016fairest}. In addition to the axiomatic approach, several other means of explanation have been proposed and analyzed, such as supplying counterfactuals \cite{mothilal2020explaining, lucic2020does}, dialogue-games based on argumentation \cite{sklar2018explanation}, local approximations \cite{ribeiro2016should}, feature-based explanations \cite{kleinerman2018providing, lundberg2017unified} and example-based explanations \cite{vivian2019humans}. 
All of these papers rely on a particular algorithm leading to the solution to explain the optimality of the solution, whereas we do not rely on one particular algorithm but on features of the solution to explain it. One exception is recent research by Zar et al \cite{zar2021explaining} that uses explanations as a means of providing additional information such as time and cost in the context of ridesharing and uses human feedback to pick the best explanation.  Still, their work uses rather ad-hoc specific explanations and there is no cohesive methodology suggesting how an initial set of explanations can be generated.  

\balance

Finally, in the context of ranked voting that is used in our experiments as the application domain, there is much we can learn from insights obtained in prior work related to human behavior. This includes a higher rate of manipulation for the Plurality voting rule \cite{bassi2015voting, forsythe1996experimental}, bias towards the candidate with the highest first place votes \cite{tal2015study} and a higher familiarity of the Plurality voting rule \cite{mccune2019can}. Surveys to analyze the understanding of ranked choice voting have reported that the respondents do not find it as difficult to follow the rules of ranked choice voting as it was perceived by the researchers \cite{nielson2017ranked, donovan2019self}.
Naturally, all of these insights can be highly valuable in constructing effective explanations but they are very particular to the domain of ranked voting while our feature-based explanations are general and can be used for any domain.

\section{Conclusion and Future Work}

Applying mechanism design and social choice theory in real life is highly challenging. Even if one manages to aggregate or combine the individual preferences into a collective social welfare function, there is no guarantee concerning people's satisfaction from and acceptance of the outcome, as these are affected by various factors. One possible method to increase satisfaction and acceptance is adapting the mechanism itself, e.g., by gaining a better understanding of how these measures are affected by the specific instance properties and the elected choice. In this paper we present a complementary approach, which uses explanations as a means for improving participant satisfaction and acceptance, for cases where the mechanism's outcome is not the most preferred by the individual. 

As we argue throughout the paper, explaining a social choice is different from classical XAI. While the latter aims to explain the optimality of the decision made for the specific user, the social-choice explanation is trying to explain the decision made in the context of a social setting. Hence, also the difference between acceptance and satisfaction. As discussed in the related work section, despite the importance of this field, most work in the area of XAI has focused, to date, on generating explanations of the first type above.

Our results show that explaining a social choice in the context of the specific social setting can have a substantial (positive) influence over individuals' satisfaction from and acceptance of the outcome picked. In particular, the extent of improvement achieved with explanations increases as we explain the selection of less favorable outcomes, based on the participant's preference index.

While the success of crowdsourced explanations was rather expected, as people are likely to be influenced by explanations made by people, we are highly encouraged by the performance of our automated explanation generation method based on theoretical features. Not only did the feature-based explanations perform as well as with crowdsourced explanations, it also managed to significantly reduce the percentage of participants who believe another candidate is more deserving of winning, compared to the latter. Furthermore, our method manages to reach such a performance even though it is inherently limited to providing only explanations related to features in which the outcome to be explained is ranked first. Meaning that for some instances it only provides a single explanation, as opposed to three explanations provided by people. All in all, the ability to automatically produce effective explanations saves substantial resources as it does not require hiring people for providing explanations from scratch any time a new instance needs to be explained. Obviously, upon accumulating substantial training data, one can develop machine learning-based prediction models in order to facilitate the extraction of an effective set of explanations. Still, much human labor is required in order to generate such data. Our automatic method does not require any human-in-the-loop intervention.

We note that even though the utmost care was taken in our experiments in order to replicate real-life settings, such as the notion of utility through an allotted order of preference, voting does not happen in a vacuum and voters are subject to external influences such as poll predictions. Hence, a prominent direction for future research is an experimental design tending to all of these issues. Another related direction is employing similar ideas and proposing algorithms to generate explanations in other domains.

Finally, we propose to extend the set of features based on which explanations can be generated. We have used six features to frame the explanations, however the literature is replete with axioms and criteria that can be considered for explanations. In the future, this research can be complemented by studies that use as many features as possible to frame the explanations while using crowdsourcing coupled with machine learning to choose the best explanations \cite{kraus2020ai}.

{\bf Acknowledgements:} This research has been partly supported by the Israeli Ministry of Science \& Technology (grants No. 89583), Israel Science Foundation under grant 1958/20, EU Project TAILOR under grant 952215 and by the Data Science Institute at Bar-Ilan University.


\bibliographystyle{ieeetr} 
\bibliography{sample}

\begin{thebibliography}{10}

\bibitem{wolfe2014driving}
S.~Wolfe, ``Driving into the ocean and 8 other spectacular fails as gps turns
  25,'' {\em Public Radio International}, 2014.

\bibitem{zahedi2020not}
Z.~Zahedi, S.~Sengupta, and S.~Kambhampati, ``Why didn't you allocate this task
  to them?'negotiation-aware task allocation and contrastive explanation
  generation,'' {\em arXiv preprint arXiv:2002.01640}, 2020.

\bibitem{donnelly2010devil}
C.~Donnelly and P.~Embrechts, ``The devil is in the tails: actuarial
  mathematics and the subprime mortgage crisis,'' {\em ASTIN Bulletin: The
  Journal of the IAA}, vol.~40, no.~1, pp.~1--33, 2010.

\bibitem{lucic2020does}
A.~Lucic, H.~Haned, and M.~de~Rijke, ``Why does my model fail? contrastive
  local explanations for retail forecasting,'' in {\em Proceedings of the 2020
  Conference on Fairness, Accountability, and Transparency}, pp.~90--98, 2020.

\bibitem{ray2019can}
A.~Ray, Y.~Yao, R.~Kumar, A.~Divakaran, and G.~Burachas, ``Can you explain
  that? lucid explanations help human-ai collaborative image retrieval,'' in
  {\em Proceedings of the AAAI Conference on Human Computation and
  Crowdsourcing}, pp.~153--161, 2019.

\bibitem{kleinerman2018providing}
A.~Kleinerman, A.~Rosenfeld, and S.~Kraus, ``Providing explanations for
  recommendations in reciprocal environments,'' in {\em Proceedings of the 12th
  ACM conference on recommender systems}, pp.~22--30, 2018.

\bibitem{kouki2019personalized}
P.~Kouki, J.~Schaffer, J.~Pujara, J.~O'Donovan, and L.~Getoor, ``Personalized
  explanations for hybrid recommender systems,'' in {\em Proceedings of the
  24th International Conference on Intelligent User Interfaces}, pp.~379--390,
  2019.

\bibitem{kraus2020ai}
S.~Kraus, A.~Azaria, J.~Fiosina, M.~Greve, N.~Hazon, L.~Kolbe, T.-B. Lembcke,
  J.~P. Muller, S.~Schleibaum, and M.~Vollrath, ``Ai for explaining decisions
  in multi-agent environments,'' in {\em Proceedings of the AAAI Conference on
  Artificial Intelligence}, pp.~13534--13538, 2020.

\bibitem{kulesza2015principles}
T.~Kulesza, M.~Burnett, W.-K. Wong, and S.~Stumpf, ``Principles of explanatory
  debugging to personalize interactive machine learning,'' in {\em Proceedings
  of the 20th international conference on intelligent user interfaces},
  pp.~126--137, 2015.

\bibitem{smith2020no}
A.~Smith-Renner, R.~Fan, M.~Birchfield, T.~Wu, J.~Boyd-Graber, D.~S. Weld, and
  L.~Findlater, ``No explainability without accountability: An empirical study
  of explanations and feedback in interactive ml,'' in {\em Proceedings of the
  2020 CHI Conference on Human Factors in Computing Systems}, pp.~1--13, 2020.

\bibitem{zar2021explaining}
D.~Zar, N.~Hazon, and A.~Azaria, ``Explaining ridesharing: selection of
  explanations for increasing user satisfaction,'' in {\em European Conference
  on Multi-Agent Systems}, pp.~89--107, Springer, 2021.

\bibitem{arrow1951social}
K.~J. Arrow, {\em Social choice and individual values}.
\newblock Yale university press, 1951.

\bibitem{brams2012maxsum}
S.~Brams, M.~Feldman, J.~Lai, J.~Morgenstern, and A.~Procaccia, ``On maxsum
  fair cake divisions,'' in {\em Proceedings of the AAAI Conference on
  Artificial Intelligence}, 2012.

\bibitem{ianovski2012cake}
E.~Ianovski, ``Cake cutting mechanisms,'' {\em arXiv preprint arXiv:1203.0100},
  2012.

\bibitem{procaccia2018fair}
A.~D. Procaccia, R.~A. Velez, and D.~Yu, ``Fair rent division on a budget,'' in
  {\em Proceedings of the Thirty-Second AAAI Conference on Artificial
  Intelligence and Thirtieth Innovative Applications of Artificial Intelligence
  Conference and Eighth AAAI Symposium on Educational Advances in Artificial
  Intelligence}, pp.~1177--1184, 2018.

\bibitem{andersson2014budget}
T.~Andersson, L.~Ehlers, and L.-G. Svensson, ``Budget balance, fairness, and
  minimal manipulability,'' {\em Theoretical Economics}, vol.~9, no.~3,
  pp.~753--777, 2014.

\bibitem{procaccia2019axioms}
A.~D. Procaccia, ``Axioms should explain solutions,'' in {\em The Future of
  Economic Design}, pp.~195--199, Springer, 2019.

\bibitem{hazon2012evaluation}
N.~Hazon, Y.~Aumann, S.~Kraus, and M.~Wooldridge, ``On the evaluation of
  election outcomes under uncertainty,'' {\em Artificial Intelligence},
  vol.~189, pp.~1--18, 2012.

\bibitem{emerson2013original}
P.~Emerson, ``The original borda count and partial voting,'' {\em Social Choice
  and Welfare}, vol.~40, no.~2, pp.~353--358, 2013.

\bibitem{young1988condorcet}
H.~P. Young, ``Condorcet's theory of voting,'' {\em American Political science
  review}, vol.~82, no.~4, pp.~1231--1244, 1988.

\bibitem{mccune2019can}
D.~McCune and L.~McCune, ``“how can we compare different voting methods?” a
  voting theory project,'' {\em PRIMUS}, vol.~29, no.~5, pp.~487--501, 2019.

\bibitem{felsenthal2018voting}
D.~S. Felsenthal and H.~Nurmi, ``Voting procedures designed to elect a single
  candidate,'' in {\em Voting Procedures for Electing a Single Candidate},
  pp.~15--25, Springer, 2018.

\bibitem{mitchell2001clustering}
J.~Mitchell, ``Clustering and psychological barriers: The importance of
  numbers,'' {\em Journal of Futures Markets: Futures, Options, and Other
  Derivative Products}, vol.~21, no.~5, pp.~395--428, 2001.

\bibitem{fehr2000fairness}
E.~Fehr and S.~G{\"a}chter, ``Fairness and retaliation: The economics of
  reciprocity,'' {\em Journal of economic perspectives}, vol.~14, no.~3,
  pp.~159--181, 2000.

\bibitem{pu2007trust}
P.~Pu and L.~Chen, ``Trust-inspiring explanation interfaces for recommender
  systems,'' {\em Knowledge-Based Systems}, vol.~20, no.~6, pp.~542--556, 2007.

\bibitem{10.1145/1978942.1978963}
J.~O. Wobbrock, L.~Findlater, D.~Gergle, and J.~J. Higgins, {\em The Aligned
  Rank Transform for Nonparametric Factorial Analyses Using Only Anova
  Procedures}, p.~143–146.
\newblock New York, NY, USA: Association for Computing Machinery, 2011.

\bibitem{bassi2015voting}
A.~Bassi, ``Voting systems and strategic manipulation: An experimental study,''
  {\em Journal of Theoretical Politics}, vol.~27, no.~1, pp.~58--85, 2015.

\bibitem{forsythe1996experimental}
R.~Forsythe, T.~Rietz, R.~Myerson, and R.~Weber, ``An experimental study of
  voting rules and polls in three-candidate elections,'' {\em International
  Journal of Game Theory}, vol.~25, no.~3, pp.~355--383, 1996.

\bibitem{tal2015study}
M.~Tal, R.~Meir, and Y.~Gal, ``A study of human behavior in online voting,'' in
  {\em Proceedings of the 2015 International Conference on Autonomous Agents
  and Multiagent Systems}, pp.~665--673, 2015.

\bibitem{goodman2017european}
B.~Goodman and S.~Flaxman, ``European union regulations on algorithmic
  decision-making and a “right to explanation”,'' {\em AI magazine},
  vol.~38, no.~3, pp.~50--57, 2017.

\bibitem{selbst2018meaningful}
A.~Selbst and J.~Powles, ``“meaningful information” and the right to
  explanation,'' in {\em Conference on Fairness, Accountability and
  Transparency}, pp.~48--48, PMLR, 2018.

\bibitem{10.2307/j.ctt13x0hch}
F.~Pasquale, {\em The Black Box Society: The Secret Algorithms That Control
  Money and Information}.
\newblock Harvard University Press, 2015.

\bibitem{doshi2017accountability}
F.~Doshi-Velez, M.~Kortz, R.~Budish, C.~Bavitz, S.~Gershman, D.~O'Brien,
  K.~Scott, S.~Schieber, J.~Waldo, D.~Weinberger, {\em et~al.},
  ``Accountability of ai under the law: The role of explanation,'' {\em arXiv
  preprint arXiv:1711.01134}, 2017.

\bibitem{rosenfeld2019explainability}
A.~Rosenfeld and A.~Richardson, ``Explainability in human--agent systems,''
  {\em Autonomous Agents and Multi-Agent Systems}, vol.~33, no.~6,
  pp.~673--705, 2019.

\bibitem{sokol2020explainability}
K.~Sokol and P.~Flach, ``Explainability fact sheets: a framework for systematic
  assessment of explainable approaches,'' in {\em Proceedings of the 2020
  Conference on Fairness, Accountability, and Transparency}, pp.~56--67, 2020.

\bibitem{hall2019systematic}
M.~Hall, D.~Harborne, R.~Tomsett, V.~Galetic, S.~Quintana-Amate, A.~Nottle, and
  A.~Preece, ``A systematic method to understand requirements for explainable
  ai (xai) systems,'' in {\em Proceedings of the IJCAI Workshop on eXplainable
  Artificial Intelligence (XAI 2019), Macau, China}, vol.~11, 2019.

\bibitem{verhagen2021two}
R.~S. Verhagen, M.~A. Neerincx, and M.~L. Tielman, ``A two-dimensional
  explanation framework to classify ai as incomprehensible, interpretable, or
  understandable,'' in {\em International Workshop on Explainable, Transparent
  Autonomous Agents and Multi-Agent Systems}, pp.~119--138, Springer, 2021.

\bibitem{schelenz2020best}
L.~Schelenz, A.~Segal, and K.~Gal, ``Best practices for transparency in machine
  generated personalization,'' in {\em Adjunct Publication of the 28th ACM
  Conference on User Modeling, Adaptation and Personalization}, pp.~23--28,
  2020.

\bibitem{miller2019explanation}
T.~Miller, ``Explanation in artificial intelligence: Insights from the social
  sciences,'' {\em Artificial intelligence}, vol.~267, pp.~1--38, 2019.

\bibitem{ciatto2020agent}
G.~Ciatto, M.~I. Schumacher, A.~Omicini, and D.~Calvaresi, ``Agent-based
  explanations in ai: Towards an abstract framework,'' in {\em International
  Workshop on Explainable, Transparent Autonomous Agents and Multi-Agent
  Systems}, pp.~3--20, Springer, 2020.

\bibitem{amgoud2009using}
L.~Amgoud and H.~Prade, ``Using arguments for making and explaining
  decisions,'' {\em Artificial Intelligence}, vol.~173, no.~3-4, pp.~413--436,
  2009.

\bibitem{sohrabi2011preferred}
S.~Sohrabi, J.~A. Baier, and S.~A. McIlraith, ``Preferred explanations: theory
  and generation via planning,'' in {\em Proceedings of the Twenty-Fifth AAAI
  Conference on Artificial Intelligence}, pp.~261--267, 2011.

\bibitem{calegari2020explainable}
R.~Calegari, A.~Omicini, and G.~Sartor, ``Explainable and ethical ai: A
  perspective on argumentation and logic programming,'' in {\em International
  Conference of the Italian Association for Artificial Intelligence},
  pp.~19--36, Springer, 2020.

\bibitem{cailloux2016arguing}
O.~Cailloux and U.~Endriss, ``Arguing about voting rules,'' in {\em Proceedings
  of the 2016 International Conference on Autonomous Agents \& Multiagent
  Systems}, pp.~287--295, 2016.

\bibitem{boixel2020automated}
A.~Boixel and U.~Endriss, ``Automated justification of collective decisions via
  constraint solving,'' in {\em Proceedings of the 19th International
  Conference on Autonomous Agents and MultiAgent Systems}, pp.~168--176, 2020.

\bibitem{belahcene2018accountable}
K.~Belahcene, Y.~Chevaleyre, C.~Labreuche, N.~Maudet, V.~Mousseau, and
  W.~Ouerdane, ``Accountable approval sorting,'' in {\em Proceedings of the
  27th International Joint Conference on Artificial Intelligence}, pp.~70--76,
  2018.

\bibitem{boixel2021complexity}
A.~Boixel and R.~de~Haan, ``On the complexity of finding justifications for
  collective decisions,'' in {\em Proceedings of the AAAI Conference on
  Artificial Intelligence}, pp.~5194--5201, 2021.

\bibitem{NEURIPS2020_10c72a9d}
D.~Peters, A.~D. Procaccia, A.~Psomas, and Z.~Zhou, ``Explainable voting,'' in
  {\em Advances in Neural Information Processing Systems}, vol.~33,
  pp.~1525--1534, Curran Associates, Inc., 2020.

\bibitem{gal2016fairest}
Y.~Gal, M.~Mash, A.~D. Procaccia, and Y.~Zick, ``Which is the fairest (rent
  division) of them all?,'' in {\em Proceedings of the 2016 ACM Conference on
  Economics and Computation}, pp.~67--84, 2016.

\bibitem{mothilal2020explaining}
R.~K. Mothilal, A.~Sharma, and C.~Tan, ``Explaining machine learning
  classifiers through diverse counterfactual explanations,'' in {\em
  Proceedings of the 2020 Conference on Fairness, Accountability, and
  Transparency}, FAT* '20, (New York, NY, USA), p.~607–617, Association for
  Computing Machinery, 2020.

\bibitem{sklar2018explanation}
E.~I. Sklar and M.~Q. Azhar, ``Explanation through argumentation,'' in {\em
  Proceedings of the 6th International Conference on Human-Agent Interaction},
  pp.~277--285, 2018.

\bibitem{ribeiro2016should}
M.~T. Ribeiro, S.~Singh, and C.~Guestrin, ``" why should i trust you?"
  explaining the predictions of any classifier,'' in {\em Proceedings of the
  22nd ACM SIGKDD international conference on knowledge discovery and data
  mining}, pp.~1135--1144, 2016.

\bibitem{lundberg2017unified}
S.~M. Lundberg and S.-I. Lee, ``A unified approach to interpreting model
  predictions,'' in {\em Proceedings of the 31st international conference on
  neural information processing systems}, pp.~4768--4777, 2017.

\bibitem{vivian2019humans}
V.~Lai and C.~Tan, ``On human predictions with explanations and predictions of
  machine learning models: A case study on deception detection,'' in {\em
  Proceedings of the Conference on Fairness, Accountability, and Transparency},
  FAT* '19, (New York, NY, USA), p.~29–38, Association for Computing
  Machinery, 2019.

\bibitem{nielson2017ranked}
L.~Nielson, ``Ranked choice voting and attitudes toward democracy in the united
  states: Results from a survey experiment,'' {\em Politics \& Policy},
  vol.~45, no.~4, pp.~535--570, 2017.

\bibitem{donovan2019self}
T.~Donovan, C.~Tolbert, and K.~Gracey, ``Self-reported understanding of
  ranked-choice voting,'' {\em Social Science Quarterly}, vol.~100, no.~5,
  pp.~1768--1776, 2019.

\end{thebibliography}


\end{document}